\documentclass[aps,prl,twocolumn,groupedaddress,amsmath,amssymb]{revtex4-1}

\usepackage[font=scriptsize,skip=0pt, justification=raggedright, singlelinecheck=false, format=plain]{caption}
\usepackage{subcaption}
\usepackage{float}
\usepackage{blindtext}

\usepackage[textsize=tiny]{todonotes}
\usepackage{dcolumn}
\usepackage{amsbsy}
\usepackage{amsmath}
\usepackage{url}
\usepackage{hyperref}
\usepackage{amssymb}
\usepackage{braket}
\usepackage{color}
\usepackage{graphicx}
\usepackage{bm}

\usepackage{cleveref}

\usepackage{comment}
\usepackage[normalem]{ulem}

\newcommand{\bff}{\mathbf{f}}
\newcommand{\bk}{\mathbf{k}}

\newcommand{\bu}{\mathbf{u}}

\newcommand{\bx}{\mathbf{x}}

\newcommand{\bK}{\mathbf{K}}

\newcommand{\bX}{\mathbf{X}}

\newcommand{\bsigma}{\boldsymbol{\sigma}}
\newcommand{\bkappa}{\boldsymbol{\kappa}}

\begin{document}

\title{Non-Schmid Effects In Perfect Single Crystal Metals At Zero Temperature}
\author{Hossein Salahshoor}
\affiliation{
   School of Aerospace Engineering,\\
   Georgia Institute of Technology,\\
   Atlanta, GA 
   }
\author{Raj Kumar Pal}
\affiliation{
   School of Aerospace Engineering,\\
   Georgia Institute of Technology,\\
   Atlanta, GA 
   }
   
\author{Julian J. Rimoli}
\email[]{rimoli@gatech.edu}
\affiliation{
   School of Aerospace Engineering,\\
   Georgia Institute of Technology,\\
   Atlanta, GA 
   }
\date{\today}
\begin{abstract}
A long standing postulate in crystal plasticity of metals is that yielding commences once the resolved shear stress on a slip plane reaches a critical value. This assumption, known as Schmid law, implies that the onset of plasticity is independent of the normal stress acting on the slip plane. We examine the validity of this assumption in single crystal perfect lattices at zero temperature by subjecting them to a wide range of combined normal and shear stresses and identifying the onset of plasticity. We employ phonon stability analysis on four distinct single crystal metals and identify the onset of plasticity with the onset of an instability. Our results show significant dependence of yielding on the normal stress, thereby illustrating the necessity of considering non-Schmid effects in crystal plasticity. Finally, contrary to the common assumption that instabilities in single crystals are of long wavelength type, we show that short wavelength instabilities are abundant in the nucleation of defects for a wide range of loading conditions.  
\end{abstract}

\maketitle

Schmid law in crystal plasticity states that glide on a given slip system commences when its resolved shear stress reaches a critical value~\cite{schmid1950kristallplastizitat, ito2001atomistic}. By assuming that this value is constant, it neglects any effect that the normal component of the traction acting on the slip plane could have on yield initiation. Inspired by the physics of friction, the present letter aims to scrutinize this assumption in single crystal metals at zero temperature and demonstrate how, at certain stress levels, the critical resolved shear stress depends on the normal traction component.

Loading conditions leading to the onset of plasticity have been extensively investigated over the last century. 
In 1900, Guest \cite{guest1900v} showed through various experiments that yielding starts when the maximum shear stress reaches a critical value. This idea was utilized by M.T. Huber (unpublished) and von Mises \cite{osakada2010history} to propose yield criteria. The following decade, crystal plasticity and the effect of material texture became a major research focus. G. I. Taylor, in his 1934 seminal paper, explained the basics of deformation mechanisms in crystal plasticity \cite{taylor1934mechanism}. In the same year, Boas and Schmid proposed their celebrated Schmid law in crystal plasticity. Later, Bridgman conducted several experiments on a sample under various hydrostatic pressures and concluded that the yield stress is independent of the hydrostatic pressure \cite{bridgman1946tensile}.

However, in 1983, Christian \cite{christian1983some} observed non-Schmid effects experimentally in iron and other body-centered cubic (BCC) metals. Similar behaviors were observed in certain alloys, like Ni$_3$Al \cite{paidar1984theory}, and crystal plasticity models based on non-Schmid effects have been proposed \cite{paidar1984theory, qin1992non}. These models attribute non-Schmid effects to the non-closed packedness present in BCC single crystals. Molecular dynamics simulations have also been employed to study non-Schmid effects in various materials \cite{ito2001atomistic,tschopp2007atomistic}. The aforementioned models and numerical studies were either phenomenological, like most models in crystal plasticity or based on molecular dynamics simulations at finite temperature.

The abundance of defects (e.g., dislocations) in bulk crystalline materials has resulted in research focusing primarily on defect evolution or nucleation from existing ones (e.g., Frank-Read source) as opposed to defect nucleation in a perfect lattice.  However, with recent advances in nanoscale devices, investigating defect nucleation in perfect lattices is gaining attention \cite{garg2015study, zhu2004predictive, li2002atomistic}. We investigate the validity of Schmid law by looking at the onset of plasticity, i.e. defect nucleation, in perfect single crystal metals. It is worth mentioning that, since stress is a continuum notion and this paper focuses on discrete lattices, in the remainder of the paper the various loading configurations are attained by enforcing displacement boundary conditions and computing the equivalent stresses from an energetic perspective. 

Four common metals are chosen for this study: Fe (BCC), Cu (FCC), Ag (FCC) and Ni (FCC). We deliberately chose three common FCC metals in order to test the hypothesis of non-closed packedness leading to non-Schmid effects \cite{ito2001atomistic}. Defect nucleation is identified by lattice instability analysis using phonon calculations. We model inter-atomic interactions through Mishin potentials, which belong to the widely adopted class of Embedded-Atom Method (EAM) potentials \cite{mishin2001structural,chamati2006embedded,williams2006embedded,mishin1999interatomic}.

Each lattice system is constructed using its standard primitive vectors written in the standard basis \cite{setyawan2010high}. We consider a sufficiently large lattice such that finite size effects are not encountered and subject the atoms at the boundary of the lattice to an affine deformation with deformation gradient $\bold F$. The deformed configuration of the boundary is obtained by $\bold x=\bold F \bX$, where $\bold x$ and $\bX$ are the position vectors of the atoms in the deformed and the reference configurations, respectively.

In our study, the applied deformation gradient consists of a hydrostatic component $\beta$ and a simple shear part $\gamma$, as follows
\begin{equation*}
\bold F= 
\begin{bmatrix}
1+\beta & \gamma & 0 \\
0 & 1+\beta & 0 \\
0 & 0 & 1+\beta
\end{bmatrix}.
\end{equation*}
This particular choice of $\bold F$ is motivated by our objective to investigate the effect of hydrostatic deformation $\beta$ on the shear value at the instability point, which we indicate as $\gamma_d$. Also, for the sake of consistency in both the lattice systems, $\bold F$ is written in the standard basis. An alternate approach to examine the Schmid law would consist in applying $\bold F$ in a coordinate system in which two of the basis vectors lie within one of the slip planes (for FCC). Note that this would result in the same hydrostatic strain as in our approach, as the hydrostatic component of the deformation gradient is invariant under a coordinate transformation. We believe this indirect way of subjecting the slip plane to shear loading allows us to compare the shear-normal coupling in FCC and BCC metals under identical deformation gradient.\\
The deformation gradient $\bold F$ is applied in two stages: We first subject the lattice to a hydrostatic deformation ($\beta$) and then impose shear deformation ($\gamma$). At a fixed $\beta$, shear deformation is increased from 0 in small steps until the onset of instability, at $\gamma=\gamma_d$.This procedure is performed for 80 different values of $\beta$, spanning from -0.04 to 0.04. \\
Note that under this procedure the lattice undergoes an affine deformation, and that an affine displacement for atoms is always a valid equilibrium solution for the lattice, owing to translational symmetry. As the deformation increases, this affine deformation solution becomes unstable and a defect nucleates. Indeed, since the affine deformation is stable to infinitesimal perturbations before the instability point, no defect nucleates under quasistatic loading. The nucleation of defects in a real single crystal beyond this critical deformation point will depend on finite size effects and on the nature of perturbations in the lattice. Our procedure thus seeks to identify a lower bound for the yield strength of the material under the considered boundary conditions. \\
We seek to identify the instability point by analyzing the stability under infinitesimal perturbations in the Fourier space, corresponding to a phonon stability analysis \cite{elliott2006stability, tadmor1999mixed}. Since instabilities are associated with the loss of positive definiteness of the Hessian of the energy functional, we use a second order approximation about the deformed configuration and write the equilibrium of an arbitrary atom `r' located in the interior, far from the boundary as:

\begin{equation}
\sum_{s=1}^{N} \bold K_{rs} \delta \bold u_{s}= \delta \bff_{r}, 
\label{eq:one}
\end{equation}
where N is the number of atoms in the lattice. $\bK$ and $\delta \bu$ are, respectively, the second derivative of the potential energy with respect to atomic coordinates and the change in displacement with respect to the deformed configuration due to force perturbation, $\delta \bff$. Without loss of generality, the origin is placed at the atom `r' and the discrete Fourier transform of $\delta \bu$ is written as:  \\
\begin{equation}
\delta \bu_s=\sum_{h=1}^{N} e^{-i\bk_h \cdot \bx_{s}} \delta \hat{\bu}_{h}, 
\label{eq:changebasis}
\end{equation}
where $\bk$ and $\delta \hat{\bu}$ are wave vector in the reciprocal basis and the Fourier transform of the displacement perturbation, respectively. Substituting Eq.~\ref{eq:changebasis} into Eq.~\ref{eq:one} and employing periodicity and orthogonality of the Fourier basis leads to:
\begin{equation}
\sum_{s=1}^{N} \bK_{rs} e^{-\mathit {i} \bkappa.\bx_s}\delta \hat {\bu}_{r} = \delta \hat {\bff}_{r}, 
\label{eq:equifourier}\
\end{equation}
where similarly $\delta \hat{\bff}$ is the force perturbation in the Fourier space. Eq.~\ref{eq:equifourier} is written for a fixed wave vector $\bk=\bkappa$. Lattice instability is identified by the loss of positive definiteness of the stiffness matrix $\sum_{s=1}^{N} \bK_{rs} e^{-\mathit {i} \bkappa.\bx_s}$. The analysis is performed by looking at the entire first Birillioun Zone (BZ) in the deformed configuration. 
\begin{figure}{}
\includegraphics[trim={3cm 6cm 3cm 6cm}, scale=0.45]{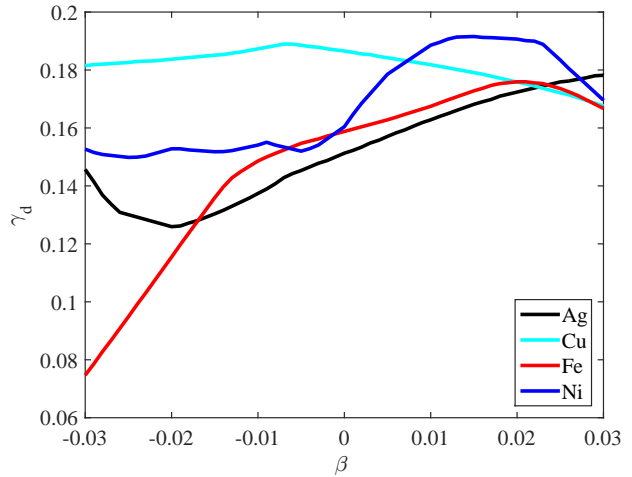}
\caption{$\gamma_d$ .vs. $\beta$: phonon stability results demonstrate a significant shear-normal coupling.}
\label{Fig:DefElastic} 
\end{figure}\\
The key message of this work is summarized in Fig.~\ref{Fig:DefElastic}, illustrating the dependence of the shear at the onset of instability, $\gamma_d$, on the hydrostatic deformation $\beta$. Since Schmid law focuses on the stress values, it is worth emphasizing that this dependency in the deformation space is an indirect way of investigating dependency in the stress space. Indeed, an independency 
of critical shear stress $\tau_c$ on hydrostatic pressure $P$ implies an independency of the critical shear strain $\gamma_d$ on the 
hydrostatic strain $\beta$. 
Note that the form of dependency $\tau_c(P)$ can be different, in general, 
from $\gamma_d(\beta)$ due to geometric and material nonlinear effects.
Schmid law implies a horizontal line, i.e. the critical shear strain $\gamma_d$ is independent of the hydrostatic deformation $\beta$. As mentioned previously, there is experimental evidence of non-Schmid effects in BCC metals, while FCC metals are considered to follow the Schmid plasticity \cite{ito2001atomistic}. Indeed, a strong coupling exists in iron (BCC). However, Fig.~\ref{Fig:DefElastic} illustrates that hydrostatic deformation significantly affects shear instabilities in FCC metals too. We observe that the trend and extent  of shear normal coupling are considerably different in various materials. While silver does not exhibit Schmid type behavior, copper and nickel follow closely the Schmid assumption in a vast regime of deformations. 

\begin{figure}{}
  \begin{subfigure}{0.45\linewidth}
    \includegraphics[trim={3cm 7cm 0cm 6cm}, scale=0.23]{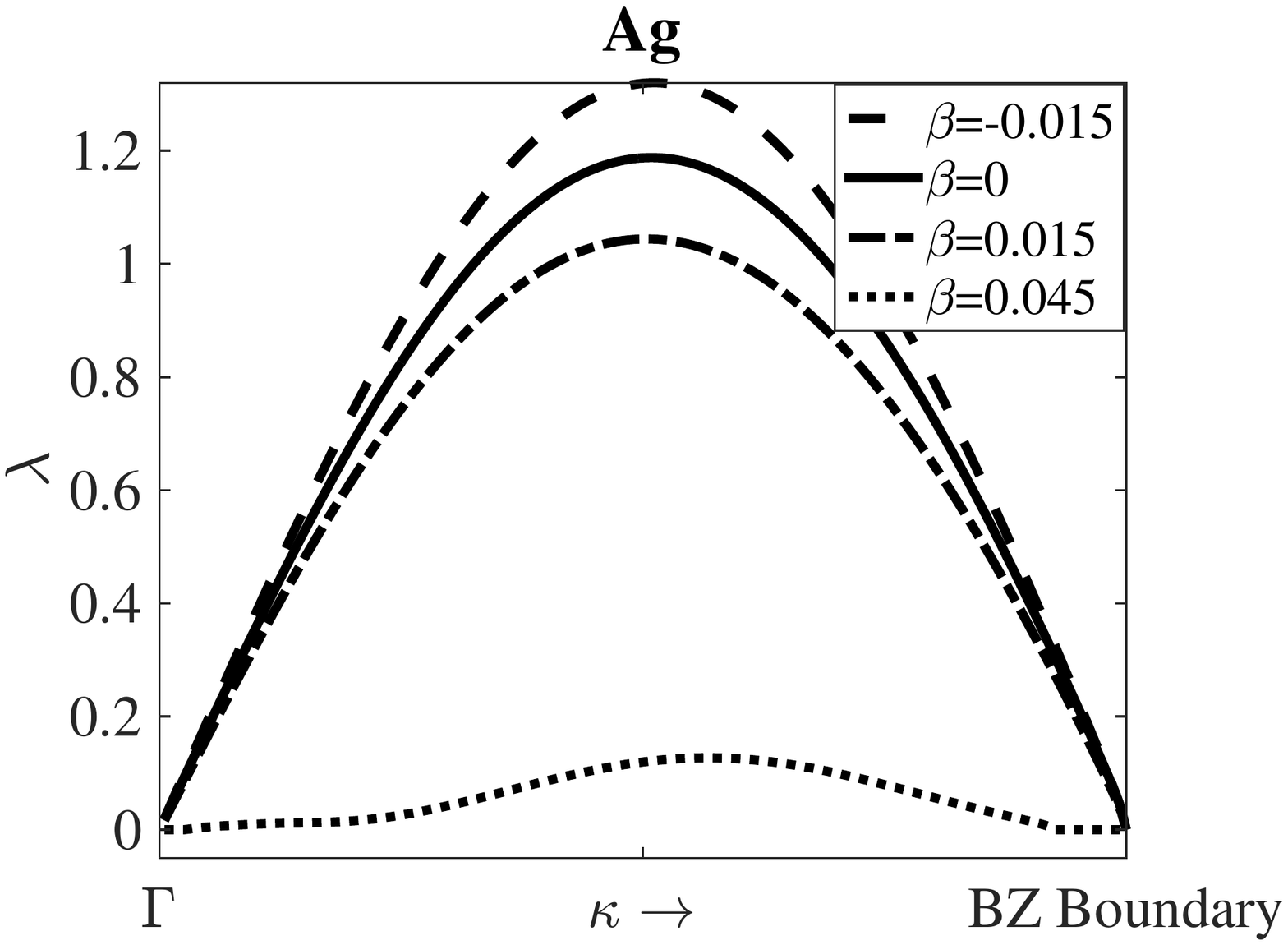}
    \centering
    \caption{} 
  \end{subfigure}%
  \begin{subfigure}{0.45\linewidth}
    \includegraphics[trim={0cm 7cm 3cm 6cm}, scale=0.23]{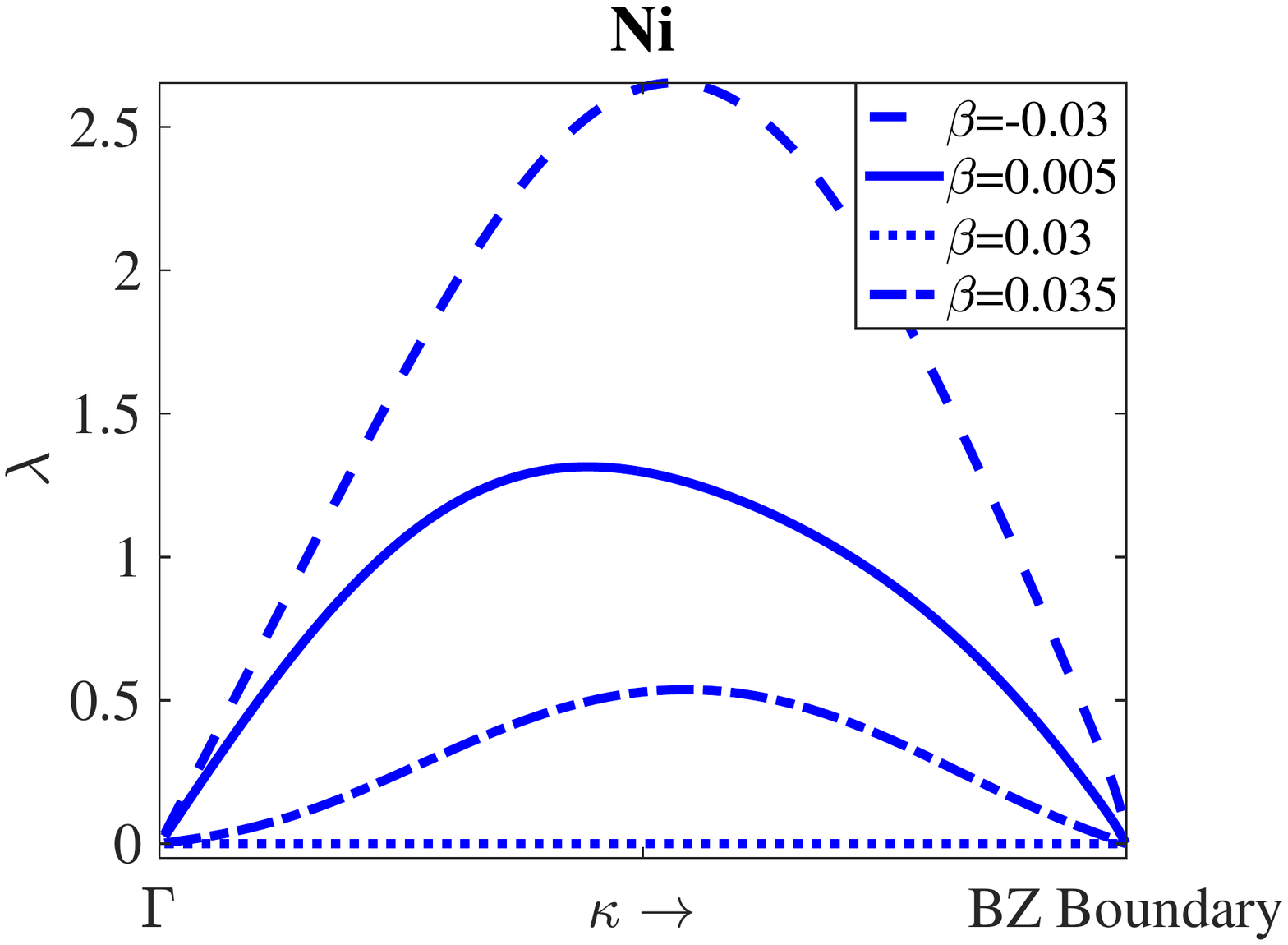}
    \centering
    \caption{}
  \end{subfigure}\par
  \begin{subfigure}[b]{0.45\linewidth}
    \includegraphics[trim={3cm 7cm 0cm 6cm}, scale=0.23]{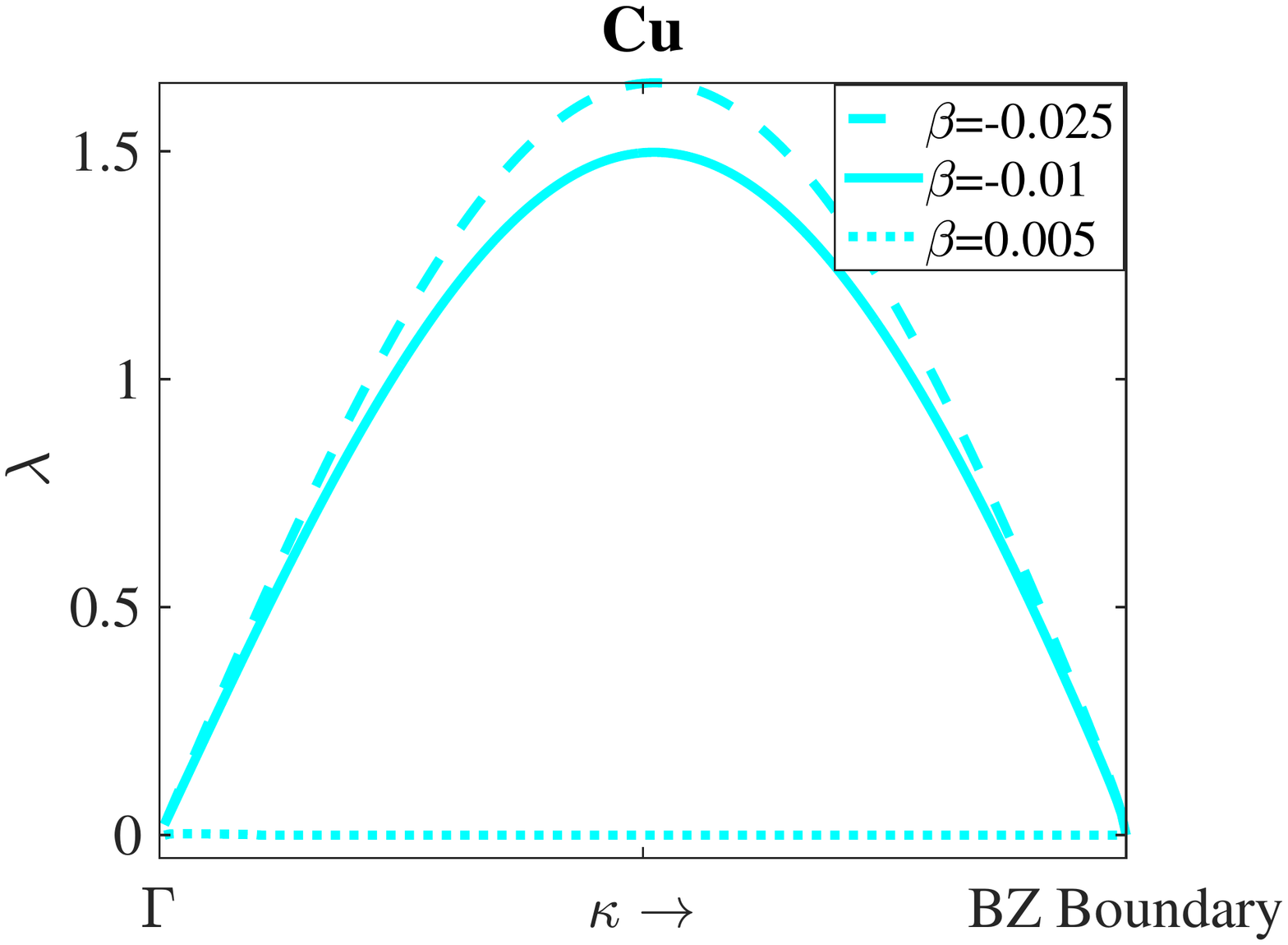}
    \centering
    \caption{}
  \end{subfigure}%
  \begin{subfigure}[b]{0.45\linewidth}
    \includegraphics[trim={0cm 7cm 3cm 6cm}, scale=0.23]{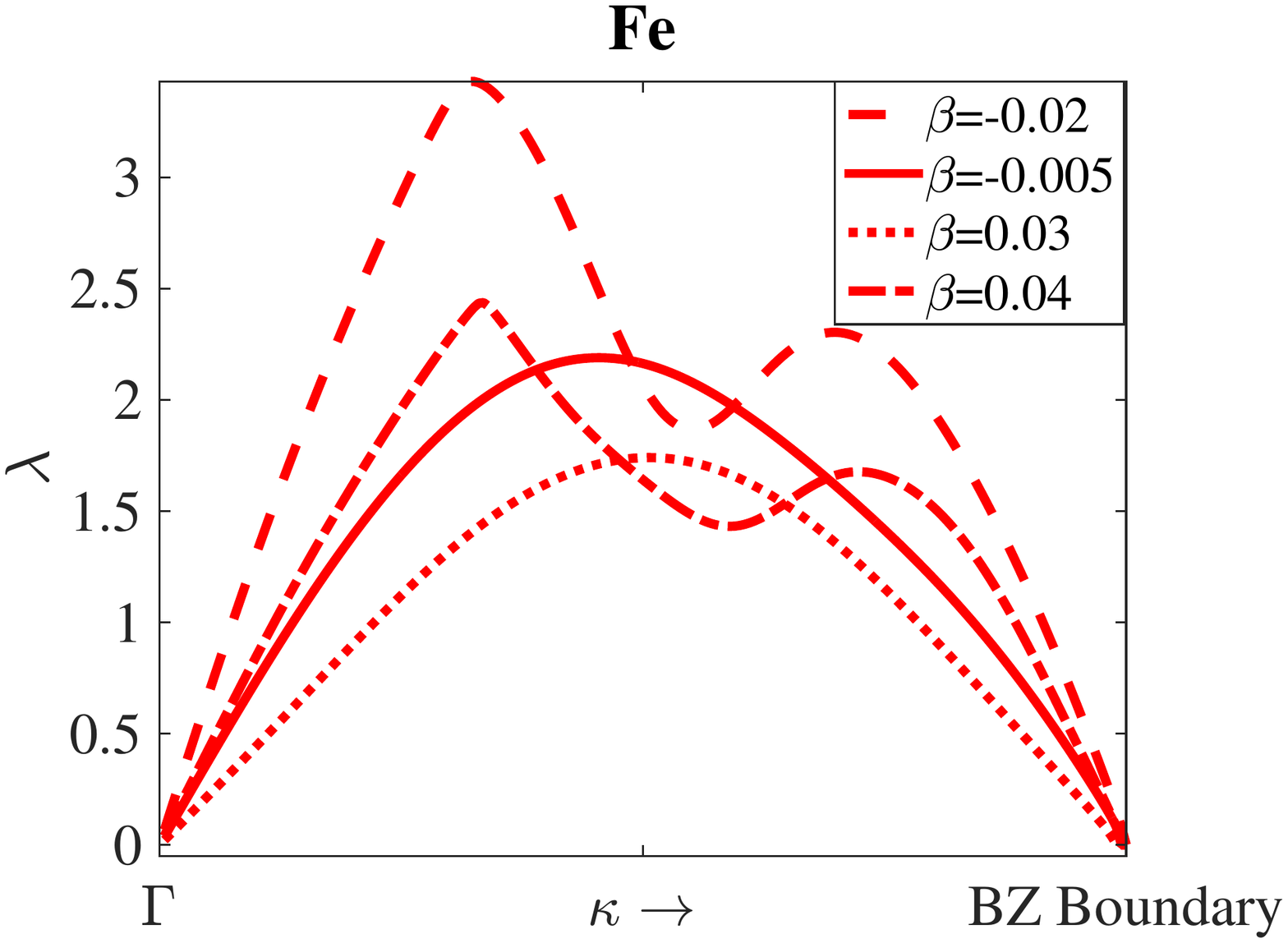}
    \centering
    \caption{}
  \end{subfigure} 
\caption{(a)-(d) $\lambda$ (square root of eigenvalues) of stiffness matrix along the path connecting the $\Gamma$ to the $\bkappa_d$: (a) Silver: For $\beta>0.04$, instabilities of long-wavelength occur. (b) Nickel: For $\beta>0.032$, instabilities of long-wavelength occur. (c) Copper: Only for $\beta<-0.006$ (hydrostatic compression), short wavelength instabilities exist. (d) Iron: All the instabilities are of short wavelength nature.}
\label{Fig:InstabilityType}
\end{figure} 
To investigate the nature of instabilities, after the wave vector associated with the instability point (i.e. $\bkappa_{d}$) is computed, we look at all the wave vectors along the line connecting $\Gamma$ to the boundary of the first BZ, passing through $\bkappa_{d}$. If the phonon softens along the entire line (eigenvalues become zero), it corresponds to a long wavelength instability. On the other hand, nonzero eigenvalues close to $\kappa=0$ and along that path imply short wavelength instability. We performed this calculation for all the loading conditions. Note that the square root of eigenvalues of the stiffness matrix, denoted by $\lambda$, is proportional to the frequency of the acoustic modes. Fig.~\ref{Fig:InstabilityType} shows $\lambda$ values for the considered materials. For each material, four different $\beta$ values are chosen and their associated $\lambda$ values are plotted along the aforementioned path.  We observe that for $\beta<0$, i.e. hydrostatic compression, instabilities of short wavelength happen in the Ag, Ni and Fe. In Ni, Cu and Ag, a transition from short to long wavelength instabilities occur. The transition point depends significantly on the material: it happens around $\beta=-0.005$ for Cu, while it only happens, at $\beta>0.04$ for Ag. To illustrate these transitions, we choose distinct $\beta$ values for various metals in Fig.~\ref{Fig:InstabilityType}. We also observe that instabilities in iron are always of a short wavelength nature (in the chosen deformation regime). It is widely assumed that instabilities in single crystals are of long wavelength nature and if a lattice is stable at long wavelengths, it will be stable at short wavelengths \cite{born1940stability, grimvall2012lattice}. While counterexamples have been shown previously \cite{wallace1965stability,liu2016crystal}, our results demonstrate an abundance of such short wavelength instabilities. Since short wavelength instabilities do not have a simple homogenized continuum analogue, 
they might not be captured by a first order continuum model. Based on our results, the common practice of using an elastic stability analysis to study nanoindentation and uniaxial tension at the nanoscale may have led to erroneous results \cite{zhu2004predictive,liu2010lattice, li2002atomistic}. We observe that the type of instability is dependent on both the material and the loading conditions.\\

\begin{figure}[ht]
\includegraphics[trim={3cm 6cm 3cm 6cm}, scale=0.45]{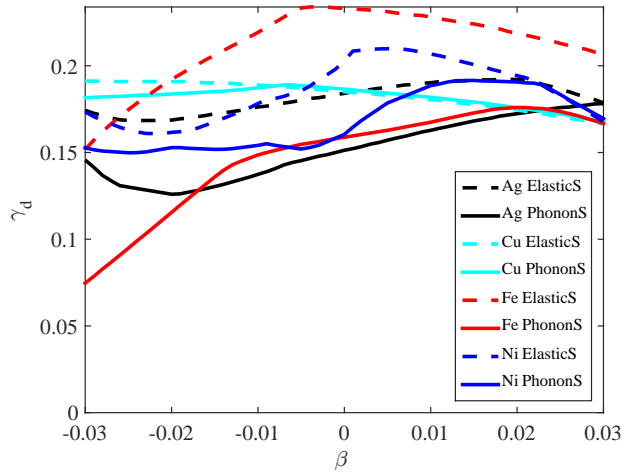}
\caption{Elastic (solid) and phonon (dashed) results: They differ, showing the occurrence of short wavelength instabilities.}
\label{Fig:DefPhonon} 
\end{figure}

To now relate the strains to the material yield stress, we consider the behavior of an equivalent continuum hyperelastic material 
until the onset of instability. We also perform an elastic stability analysis to quantify the difference between the results obtained by the two approaches
and to further illustrate the pitfalls of using an elastic stability analysis. 
For this purpose, a homogenized continuum model of the lattice is considered. Let $\overline{W}$ be the homogenized energy of the continuum under two assumptions: I) The Cauchy-Born hypothesis is satisfied, implying that the underlying lattice will deform under the same deformation gradient as the continuum, II) The strain energy density $\overline{W}(\bold F)$ in the continuum model is equal to the energy of a single unit cell normalized by its volume, obtained from the interatomic potentials \cite{pal2016continuum}. Within this framework, instability occurs following the violation of the strong ellipticity condition \cite{tadmor1999mixed, geymonat1993homogenization}:
\begin{equation}
\delta \bu^T: \frac{\partial^2 \overline{W}(\bold F)}{\partial \bold F \partial \bold F }: \delta \bu > 0, 
\label{eq:ElasticStability}
\end{equation}
where $\delta \bu$ is a perturbation of the displacement field about the deformed configuration. \\
The continuum body is subjected to exactly the same affine deformation as the lattice. We seek instabilities which result from perturbations having a plane wave basis, i.e. $\delta \bold{u} = \delta \bold{u}_0 e^{i\bkappa\cdot \bold x}$ \cite{michel2007microscopic, liu2010lattice}. Fig.~\ref{Fig:DefPhonon} compares the elastic and phonon instability results. We observe that the difference in the results are significant, e.g. 50$\%$ for iron at $\beta=-0.03$. Evidently, instabilities that happen at short or finite wavelengths are not captured by an elastic stability analysis. Indeed, since elastic instabilities under affine deformations are a subset of phonon instabilities, the $\gamma_d$ obtained by phonon calculations either coincide with elastic stability results or predict a smaller value.

\begin{figure}[ht]
\includegraphics[trim={3cm 6cm 3cm 6cm}, scale=0.45]{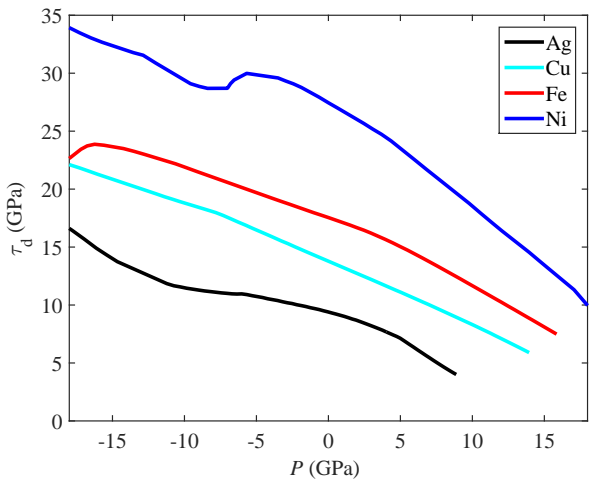}
\caption{Shear normal coupling through Cauchy stress results for the four materials: Shear .vs. hydrostatic pressure.}
\label{Fig:StressPhonon} 
\end{figure}

At the onset of instability, the Cauchy stress tensor is computed by \cite{gurtin1982introduction}:
\begin{equation}
\bsigma   = (\det \bold{F})^{-1}  \left( \dfrac{\partial W}{\partial \bold{F}}\right) \bold{F}^T.
\label{eq:Stress}
\end{equation}
Fig.~\ref{Fig:StressPhonon} illustrates the shear stress component, $\tau=\sigma_{12}$ for different hydrostatic pressure values (P), where P$=Tr(\bsigma)/3$. Evidently, by increasing the hydrostatic pressure, the shear stress at the onset of instability increases. These observations
motivate the necessity of a pressure-dependent plasticity model for perfect single crystal metals.
\begin{figure}[ht]
\includegraphics[trim={3cm 6cm 3cm 6cm}, scale=0.45]{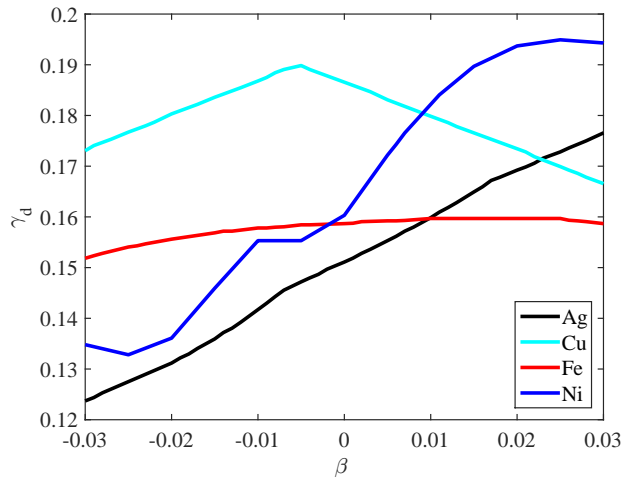}
\caption{Deformation results under combined uniaxial and simple shear deformation $\bold F'$ for the four different materials.}
\label{Fig:UniaxialDef} 
\end{figure}

While hydrostatic deformation is chosen to understand non-Schmid effects, applying this type of deformation in the nano-scale experiments might be very challenging. Fig.~\ref{Fig:UniaxialDef} is demonstrating phonon stability results for uniaxial deformation $\bold{F'}$, i.e. a combined uniaxial deformation and simple shear, which is anticipated to be useful for nano-scale experiments. 

\[
\bold{F'}= 
\begin{bmatrix}
1+\beta & \gamma & 0 \\
0 & 1 & 0 \\
0 & 0 & 1
\end{bmatrix}.
\]

In summary, the pitfalls of Schmid law are investigated and a significant dependence of the critical shear strain $\gamma_d$ on the 
hydrostatic strain $\beta$ is shown. While certain metals like copper follow the Schmid law reasonably well, others such as iron and nickel demonstrate a strong non-Schmid behavior. It is verified that depending on the crystal and the loading conditions, short wavelength instabilities could be dominant. The method pursued in this research relies on rigorous mathematical formulations. The only possible source of error is the interatomic potentials, which have been extensively  developed and used by the scientific community and are believed to be accurate enough for this study. Identifying the nature of defects, investigating the short to long wavelength transition and the role of temperature in the aforementioned phenomenon are potential future research directions.

\end{document}